\documentclass[intlimits,twoside,a4paper]{article}

\usepackage[cp1251]{inputenc}

\usepackage[eqsecnum]{cmpj3}

\usepackage{bm}
\usepackage{graphicx}

\issue{2018}{21}{4}{43702}
\doinumber{10.5488/CMP.21.43702}

\title[SNINS junction in an external magnetic field]%
{The effect of an external magnetic field on the maximum current of SNINS junctions near the critical temperature}%

\author[O.Yu. Pastukh, V.E. Sakhnyuk, A.V. Svidzinsky]%
{O.Yu. Pastukh, V.E. Sakhnyuk, A.V. Svidzinsky}
\address{Lesya Ukrainka Eastern European National University, 13 Voli Ave., 43000 Lutsk, Ukraine}

\date{Received May 24, 2018, in final form July 20, 2018}

\begin{document}

\maketitle

\begin{abstract}
The behaviour of superconducting junctions of SNINS-type (where S is superconductor, N is normal metal, I is insulator) with anharmonic current-phase relations in the external magnetic field near the critical temperature was investigated. The dependence of the maximum current on the value of the magnetic flux in a wide range of electron transmission coefficient values was considered. Also, it was investigated how the presence of a normal layer of an arbitrary thickness in the scale of coherence length and in the presence of impurities in superconducting regions affect the sensitivity of the maximum current to the magnetic field magnitude.
\keywords Josephson junctions, maximum current, magnetic flux, depairing effects
\pacs 74.50.+r
\end{abstract}

\section{Introduction}

The discovery of the Josephson effect \cite{Jos1} has greatly contributed to the development of superconductivity physics and opened a wide area for research in the field of the so-called weak superconductivity. The place where these effects exist is the Josephson junctions or any other kind of weak links \cite{Lik1,Lik2,Svi}. An important characteristic of Josephson tunnel junctions is their behaviour in the external magnetic field. As it turns out the maximum current in such structures is very sensitive to the change in the magnetic field, and for the case of a sinusoidal current-phase relation can be written as \cite{Jos2}
\begin{equation}
\nonumber
I_{\max } \left( \Phi  \right) = I_{\max } \left( 0 \right)\left| {\frac{{\sin \left( {{{\piup \Phi } \mathord{\left/{\vphantom {{\piup \Phi } {\Phi _0 }}} \right.				\kern-\nulldelimiterspace} {\Phi _0 }}} \right)}}{{{{\piup \Phi } \mathord{\left/			{\vphantom {{\piup \Phi } {\Phi _0 }}} \right.\kern-\nulldelimiterspace} {\Phi _0 }}}}} \right|,
\end{equation}
where $\Phi$ is the magnetic flux through the junction, $\Phi_0$ is the magnetic flux quantum. 
From the above expression one can see that the current drops to zero when the total magnetic flux is equal to an integer number of quantum, i.e., $\Phi  = n\Phi _0$  and can reach the values from zero to the maximum value at a change in the flow by only half the quantum $\Phi_0$. As a result, the use of this effect opens up wide opportunities for the practical application of tunnel junctions \cite{Bar,Lan,Sol,Sch}. A first experimental observation of the magnetic field dependence of the maximum Josephson current was obtained in work \cite{Row}. Lately, many experimental and theoretical investigations of this effect have been made for junctions of different types, which are classified as so-called “short” \cite{Fis,Schw1,Schw2,Bal}, with dimensions that are small compared to the Josephson penetration depth $l \ll \lambda _{\text J}$ and “long” \cite{Gol,Owe,Schw3,Ata},  for which  $l \gg \lambda _{\text J}$. Nowadays, the investigations of the behaviour of Josephson junctions in an external magnetic field for the case of annular tunnel junctions \cite{Mar,Wal} and junctions with a ferromagnetic interlayer are also actively developed \cite{Wei}.

However, a detailed analysis shows that the simple sinusoidal current-phase relation is valid only under certain conditions, which are implied on the parameters of the superconducting junction. For the case of tunnel superconducting junction of SIS-type at the temperatures close to critical, the corresponding condition is $D \ll 1$  (where $D$ is the electron transmission coefficient through the dielectric layer). At non-small values of the electron transmission coefficient, in order to obtain the correct result for the current, it is necessary to take into account the depairing effects, which under such conditions ($D \lesssim 1$) are already significant. The study of the Josephson tunnel junction, taking into account these effects, was carried out in \cite{Sak1,Bar1}, which resulted in a nontrivial result for the current-phase relation instead of a traditional sinusoidal dependence. It is obvious that the anharmonic current-phase relation will also be reflected in the behaviour of the corresponding junctions in the magnetic field. In \cite{Sak2} it is shown that for the “short” Josephson junction, an increase of electron transmission coefficient leads to an increase of the current sensitivity to a change in the external magnetic field. For the “long” SIS junction in a magnetic field \cite{Bar2}, it is shown that the intense pair breaking significantly increases the penetration depth of the magnetic field.

Nowadays, superconducting junctions of SNINS-type which combine tunnel effects and the proximity effect are of great interest both in the fundamental  and in the applied aspects of investigation. Equilibrium properties of such junctions were studied in \cite{Zai,Golb,Akh} while current-voltage characteristics of SNINS junction for various thicknesses of the normal layer both in the tunneling and in the high-transparency regime were considered in \cite{Bez,Zhi}. Our task in this work is to investigate the behaviour of superconducting junctions of SNINS-type in the external magnetic field taking into account the depairing effects. Here we consider how the presence of a wide range of electron transmission coefficient values and an arbitrary thickness of the normal layer affects the sensitivity of the maximum current to the magnitude of the magnetic field. The effect of nonmagnetic impurities is also taken  into account.

\section{Model and basic equations}

\begin{figure}[!b]
	\vspace{-2ex}%
	\centerline{\includegraphics[width=0.6\textwidth]{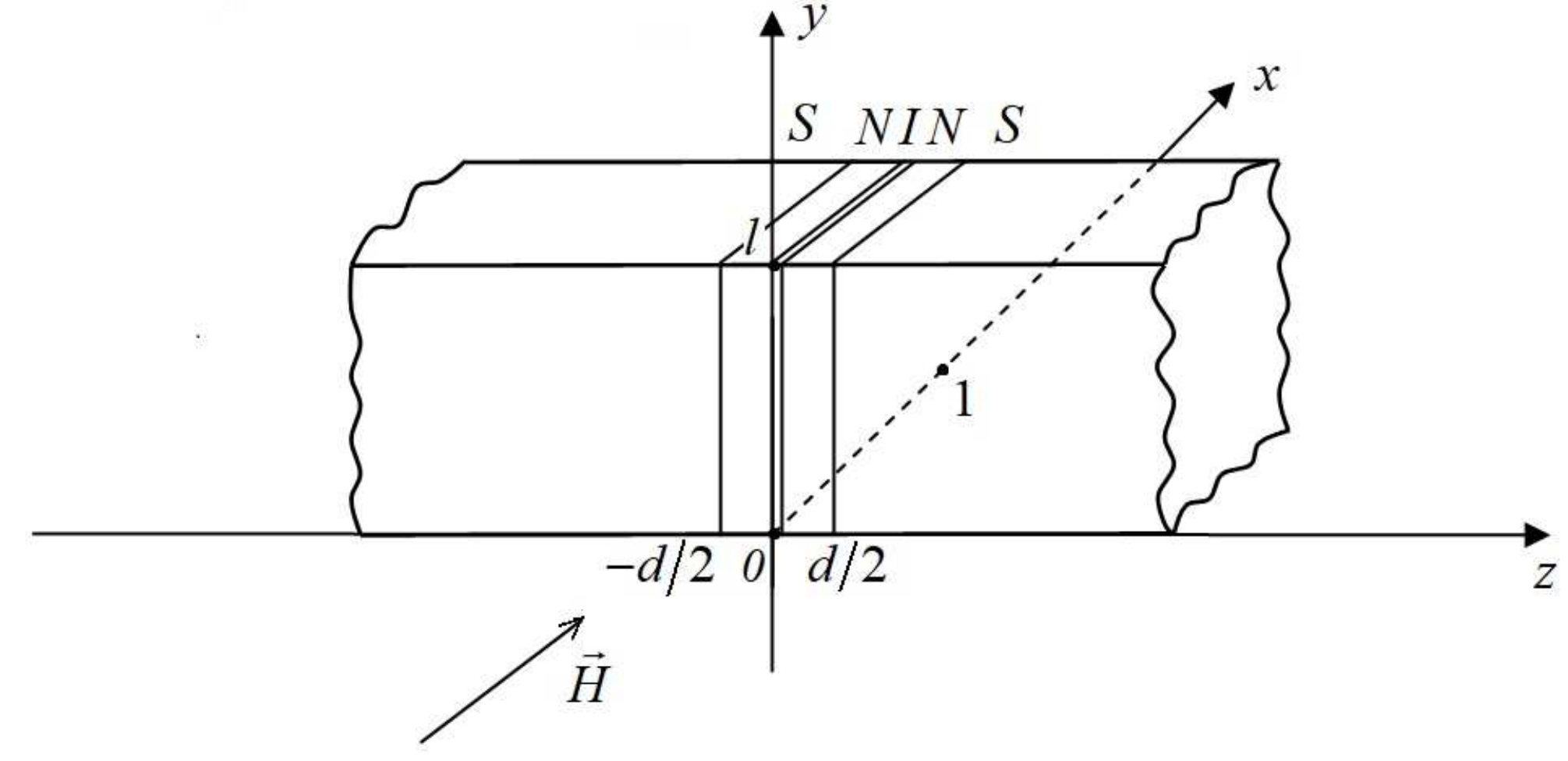}}
	\caption{Geometrical configuration of the SNINS junction.}
	\label{figure1}
\end{figure}

Let us consider the investigated superconducting junction (figure~\ref{figure1}). We choose the coordinate system so that the superconductor occupies the region $\left| z \right| > d/2$, normal metal occupies the region $\left| z \right| < d/2$, and the dielectric layer coincides with the plane $x0y$.  Let the size of the junction in the $x$-direction be equal to the unit of length and in the $y$-direction be equal to $l$.  Let the external magnetic field be applied along the $x$-axis, i.e., $\vec H = \left( {H,0,0} \right)$. The applied magnetic field is considered to be high so we can assume that the penetration depth $\lambda _{\text J}$ of the magnetic field is much greater than the size of the junction. In this case, we can neglect the self-induced magnetic field of the current passing through the junction and consider the field inside the junction as uniformly equal to the external field applied to and around the junction. For practical reasons, we take the vector potential oriented as follows: $\vec A = \left( {0, - zH,0} \right)$.
In work \cite{Pas}, where the current states in the superconducting SNINS junction near the critical temperature were investigated, it was shown that for  certain values of junction parameters, the current-phase relation is different from the traditional sinusoidal dependence. A pronounced anharmonicity of the current is caused by the depairing effects action which are significant at small values of the normal metal thickness and non-small values of the electron transmission coefficient. A significant role in this aspect also belongs to the presence of nonmagnetic impurities in the junction. That is why the behaviour of superconducting junctions with anharmonic current-phase relations in the external magnetic field becomes an interesting topic of investigation. Let us consider how the values of the electron transmission coefficient, normal metal thickness and the length of electron free path affect the dependence of the maximum current in the junction on the magnetic flux. The result for the current-phase relation in SNINS junction is given by~\cite{Pas}
\begin{equation}
\label{eqn2}
j\left( \varphi  \right) = \sin \varphi \frac{ {q_{1,\infty }  - q_{2,\infty } } }{{2\tau q_{1,\infty } q_{2,\infty } }}\frac{{\tau ^2  + \left( {\frac{{\cos ^2 \frac{\varphi }{2}}}{{q_{1,\infty }^2 }} + \frac{{\sin ^2 \frac{\varphi }{2}}}{{q_{2,\infty }^2 }}} \right)\left[ {1 - \sqrt {1 + 2\tau^2 q_{1,\infty }^2 q_{2,\infty }^2 \left( {\frac{{q_{1,\infty } \tan^2 \frac{\varphi }{2} + q_{2,\infty } }}{{q_{1,\infty }^2 \tan^2 \frac{\varphi }{2} + q_{2,\infty }^2 }}} \right)^2 } } \right]}}{{\tau^2  + \frac{1}{2}\left( {\frac{1}{{q_{2,\infty } }} - \frac{1}{{q_{1,\infty } }}} \right)^2 \sin ^2 \varphi }}\,,
\end{equation}
where $\tau=\frac{\xi_0}{\xi\left(T\right)}=[\frac{12}{7\zeta\left(3\right)}\frac{1}{\chi\left(1/\lambda\right)}(1-\frac{T}{T_{\text c}})]^{1/2}$ is the ratio of the coherence length and the characteristic length in the Ginzburg-Landau theory, $\chi ( {\frac{1}{\lambda }} ) = \frac{8}{{7\zeta \left( 3 \right)}}\sum\nolimits_{n = 0}^\infty  {\frac{1}{{\left( {2n + 1} \right)^2 \left( {2n + 1 + {1 \mathord{\left/{\vphantom {1 \lambda }} \right.		\kern-\nulldelimiterspace} \lambda }} \right)}}} $, $\lambda  = l/{\xi _0 }$ is a dimensionless electron mean free path which determines a disorder of the superconductor with nonmagnetic impurities. Coefficients $q_{1,\infty }$
and $q_{2,\infty }$ are given as follows:
\begin{align}
\nonumber
q_{1,\infty } &= \frac{1}
{{2I_1 }}\left\{ {I_2  + I_2 (a) + \frac{{\left[ {I_1  + I_1 (a)} \right]^2 }}
	{{I_0  - I_0 (a)}}} \right\},\\
\nonumber
q_{2,\infty } &= \frac{1}
{{2I_1 }}\left\{ {I_2  + I_2 (a,D) + \frac{{\left[ {I_1  + I_1 (a,D)} \right]^2 }}
	{{I_0  - I_0 (a,D)}}} \right\}.
\end{align}
In the above expressions, the notations for integrals which are determined by the kernels of linear integral equations for the order parameter were used (see \cite{Pas}).

In the presence of an external magnetic field, the phase difference between two superconductors varies in the y direction and, as a result, there appears a component of the current in the direction of the axis $0y$. The dependence of the phase difference on the coordinates can be obtained from the relation between the superfluid velocity  $\vec \upsilon _{\text s}$ and the magnetic field $\vec H$ \cite{Svi}
\begin{equation}
\nonumber
\int {\vec \upsilon _{\text s} } \rd\vec l =  - \frac{e}
{m}\int {\vec H\rd\vec S} .
\end{equation}
As an integration area, we consider a parallelepiped, where one of its sides parallel to the axis  $0y$ has an infinitely small length $\rd y$  and passes near the surface $z=d/2+0$.  The opposite side is located at the depth of the superconductor (we can assume $\vec \upsilon _{\text s}=0$  and $\vec H=0$ on it). The other two sides are parallel to the axis $0z$, and the unit vector normal to the surface is in the direction of the axis $0x$.  After the integration we choose in a similar way the integration area in the left-hand superconductor, and then we can find a jump of superfluid velocity through the junction. On the other hand, from the definition $\vec \upsilon _{\text s}  = \frac{1}{{2m}}\nabla \varphi  - \frac{e}{m}\vec A$  we can find a jump of superfluid velocity in terms of phase difference and vector potential. Thus, the expression that in a differential form determines the dependence of the coherent phase difference on the coordinates in the presence of a magnetic field is as follows:
\begin{equation}
\label{eqn3}
\frac{{\partial \varphi }}
{{\partial y}} =  - 2eH\left( {2\delta  + d} \right),
\end{equation}
where $e$  is the electron charge, $\delta$ is the magnetic field penetration depth. Integrating relation (\ref{eqn3}) and taking into account the direction of the applied magnetic field, we get
\begin{equation}
\nonumber
\varphi \left( y \right) =  - 2eH\left( {2\delta  + d} \right)y + \varphi _0 .
\end{equation}
Substituting the above expression for the phase difference into (\ref{eqn2}), one can obtain an expression for the current density $j = j\left( {H,y,\varphi _0 } \right)$, taking into account the presence of an external magnetic field. In order to find the total current through the superconducting junction, it is necessary to integrate the current density over the entire area of the junction, taking into account its size. As a result, for the total current as a function of the phase $\varphi _0$  and magnetic field $H$, we have
\begin{equation}
\label{eqn4}
I\left( {H,\varphi _0 } \right) = \int_0^1 {\rd x} \int_0^l {\rd y\,} j\left( {H,y,\varphi _0 } \right).
\end{equation}

\begin{figure}[!t]
	{\label{fig2a}\includegraphics[width=0.49\textwidth]{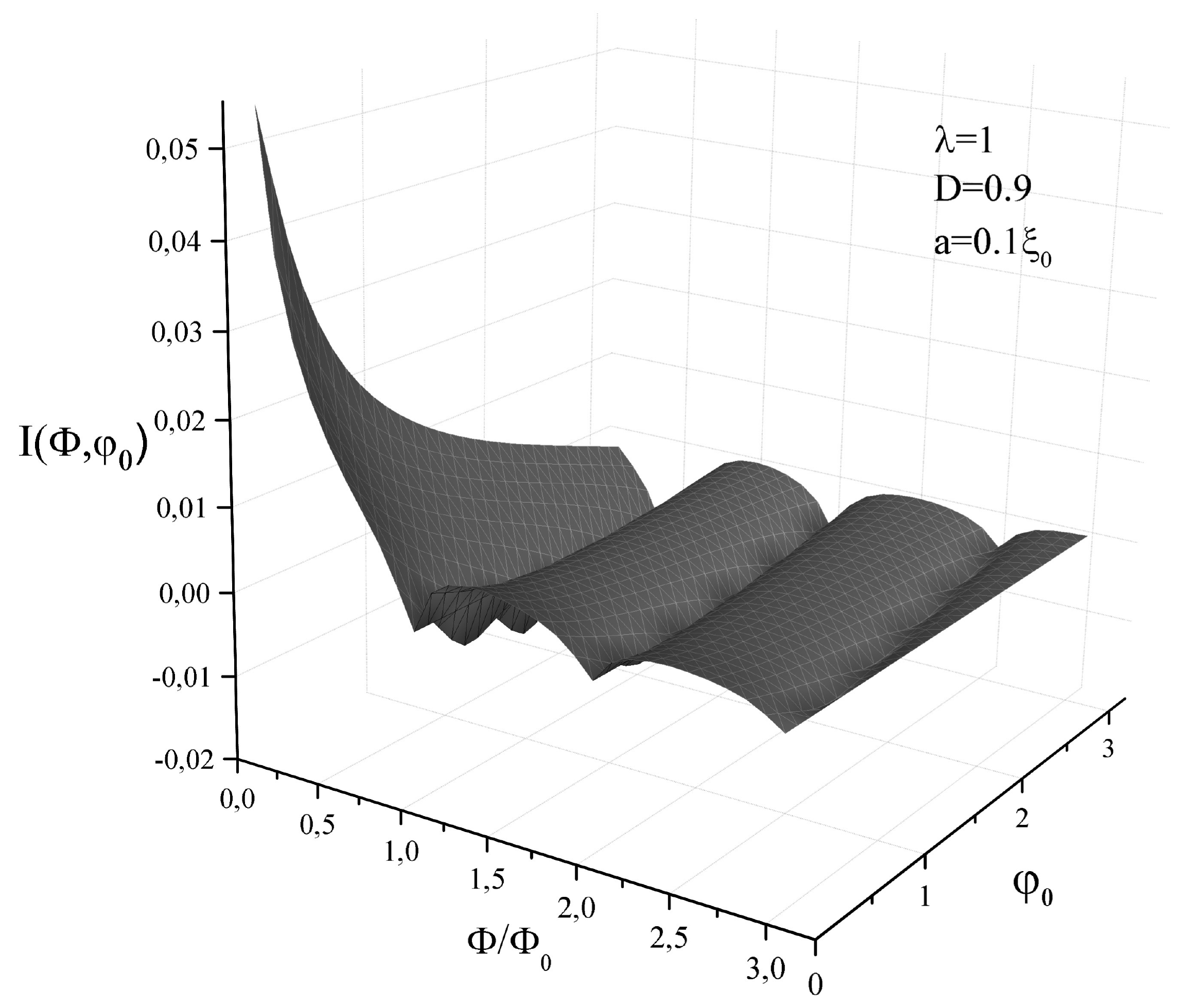}}
	{\label{fig2b}\includegraphics[width=0.49\textwidth]{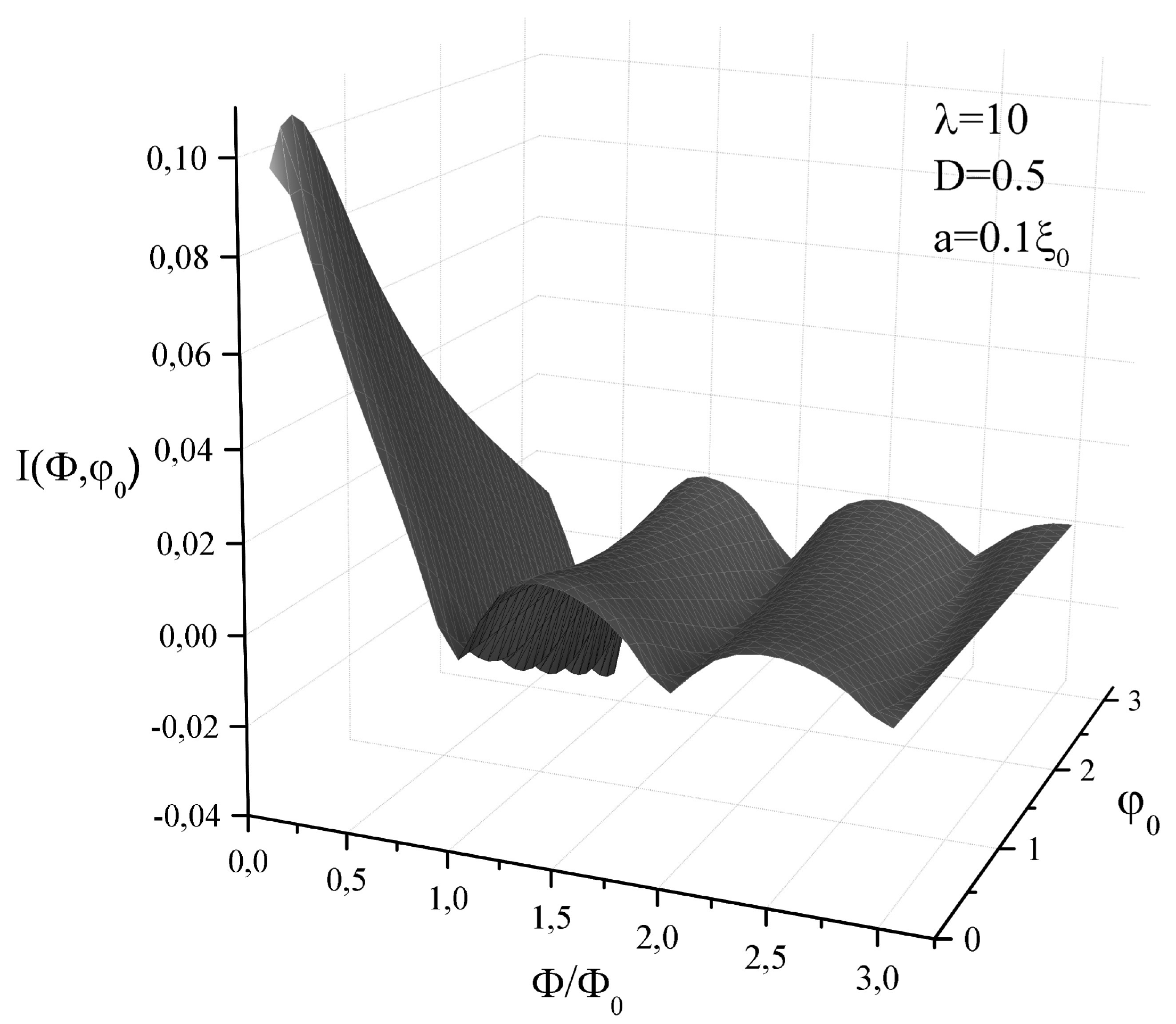}}
	\caption{Total current dependence of the magnetic flux and phase difference in SNINS junction.}
	\label{fig2}
\end{figure}

Numerical integration of (\ref{eqn4}) makes it possible to analyze the effect of the junction parameters on the dependence of the current on the magnetic flux. The corresponding graphical dependence of the total current on the magnetic field and the phase difference is shown in figure~\ref{fig2}, where the magnetic field~$H$ is expressed through the total magnetic flux $\Phi  = Hl\left( {2\delta  + d} \right)$ and the notation for the magnetic flux quantum $\Phi_0=\piup/e$ is used.

\begin{figure}[!b]
	{\label{fig3a}\includegraphics[width=0.49\textwidth]{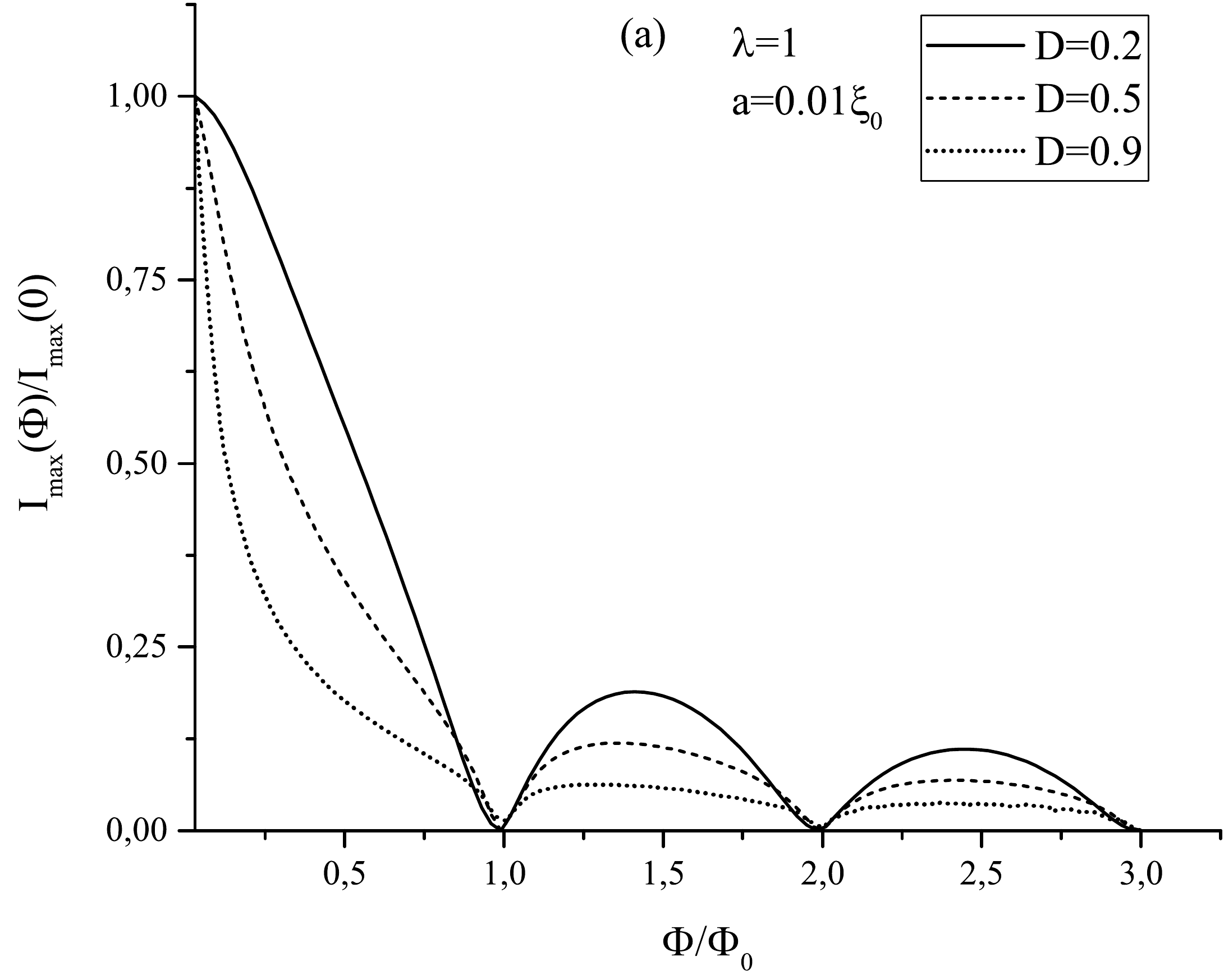}}
	{\label{fig3b}\includegraphics[width=0.49\textwidth]{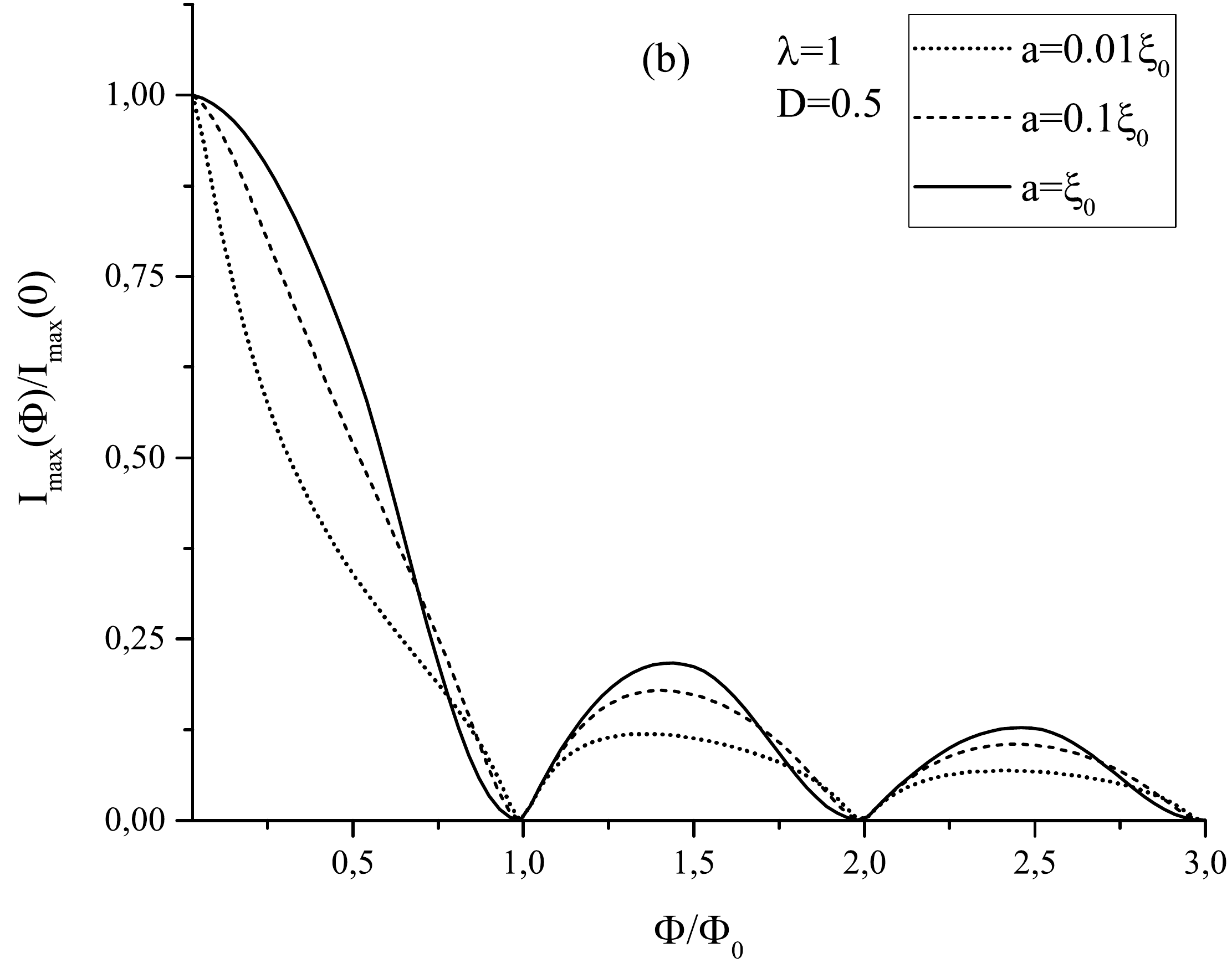}}
	\caption{Magnetic field dependence of the maximum current (normalized to the maximum current in the absence of a field) for SNINS junction.}
	\label{fig3}
\end{figure}

As we can see in figure~\ref{fig2}, with a change in the magnetic flux, there is an alternation of the maxima and minima of the current, and the values of the maxima decrease with an increase of the magnetic flux. For flux values $\Phi$ equal to an integer number of fluxons $\Phi_0$,  maximum current reduces to zero, which corresponds to the result of the classical work \cite{Row,Bal}. Comparing the figures shown above, it becomes obvious that with the decrease of the electron transmission coefficient, the peaks of the current become sharper, and the decrease in the electron free path leads to the decrease in their magnitude.

Let us now analyze how the maximum current that can be born by a short SNINS junction depends on the magnetic field. Obviously, in order to find it, it is necessary to equate the phase $\varphi _0$ derivative of $I\left( {H,\varphi _0 } \right)$ to zero, that is, to find the critical value of the phase corresponding to the maximum current. By making the calculations, one can obtain the graphical dependence of the maximum current in the junction as a function of the magnetic flux. In figure~\ref{fig3}~(a) the dependence of the maximum current as a function of the magnetic flux is shown for the given value of the electron free path $\lambda$, normal layer thickness $a$ but different values of electron transmission coefficient $D$.
An examination of the figure shows that with an increase of $D$, the maximum current in the junction becomes much more sensitive to the change in the magnetic flux, which is especially noticeable at the initial values of the magnetic flux.
This phenomenon was shown in \cite{Sak2}, where the SIS junction was considered. The dependence of the maximum current on the magnetic flux for different values of the normal layer thickness $a$ with a fixed value of the electron transmission coefficient $D$ and electron free path $\lambda$ is illustrated in figure~\ref{fig3}~(b). As one can see, the presence of the normal layer also affects the sensitivity of the current to the change in the magnetic flux --- with a decrease of the normal layer thickness sensitivity of $I_{\max } (\Phi )$ increases. From the numerical analysis for $D = 0.9$ and $\Phi / \Phi_0 = 0.3$, we find out that the maximum current of the junction decreases by 4 times, and for $D = 0.2$ and the same magnetic flux value, this decrease is only $25\%$. As a result, using in the measuring technique of superconducting SNINS junctions with non-small values of the electron transmission coefficient opens up new horizons in various fields of science. From the analysis of figure~\ref{fig3}~(b), one can conclude that with an increase of the normal layer thickness, the maximum current value $I_{\max } (\Phi )$ begins to approach the value of the maximum current in the absence of an external magnetic field $I_{\max } (0)$. Therefore, the corresponding graphs are higher. Let us now analyze how the presence of non-magnetic impurities affects the magnitude of the maximum current.
\begin{figure}[!t]
	{\label{fig4a}\includegraphics[width=0.49\textwidth]{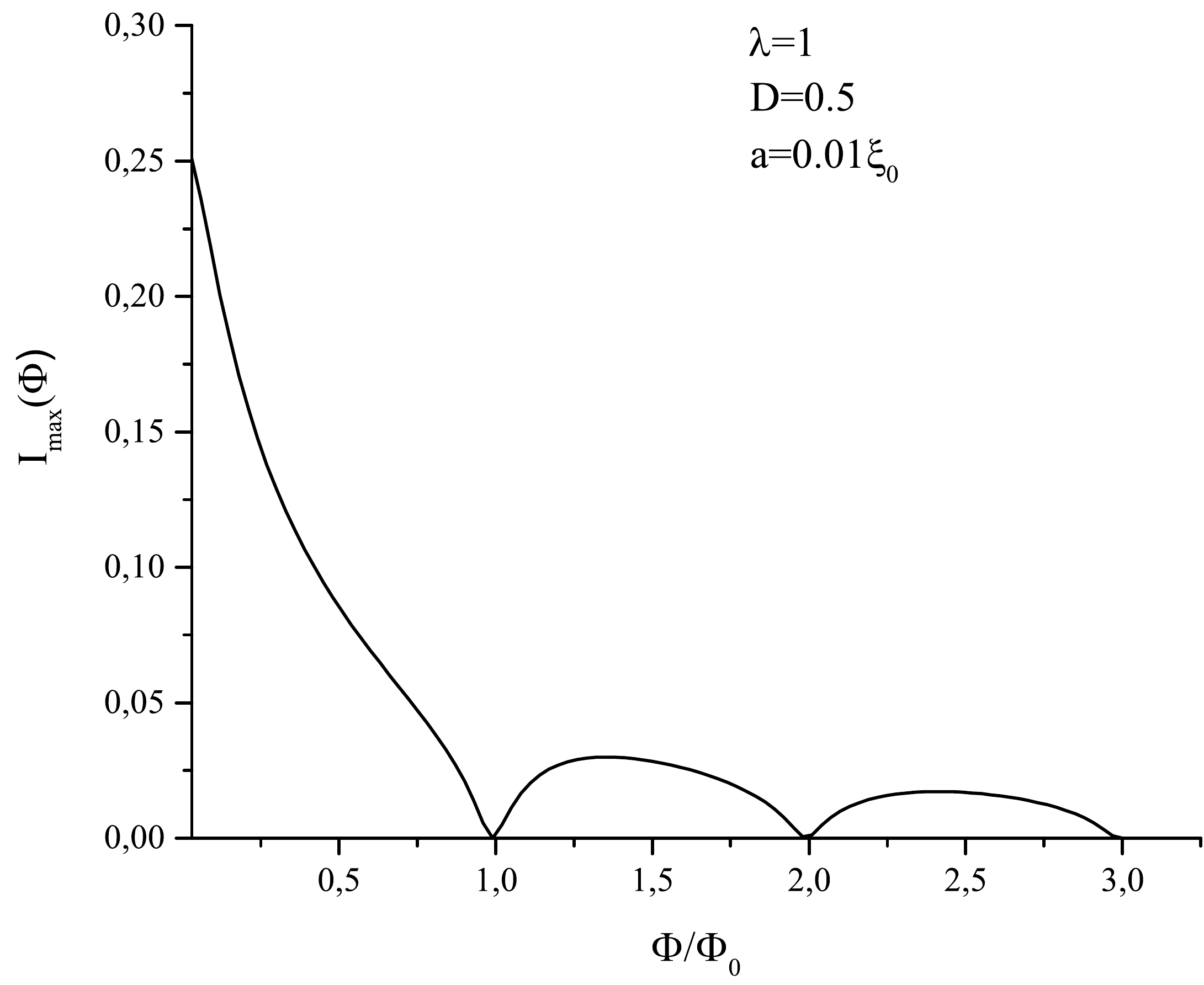}}
	{\label{fig4b}\includegraphics[width=0.49\textwidth]{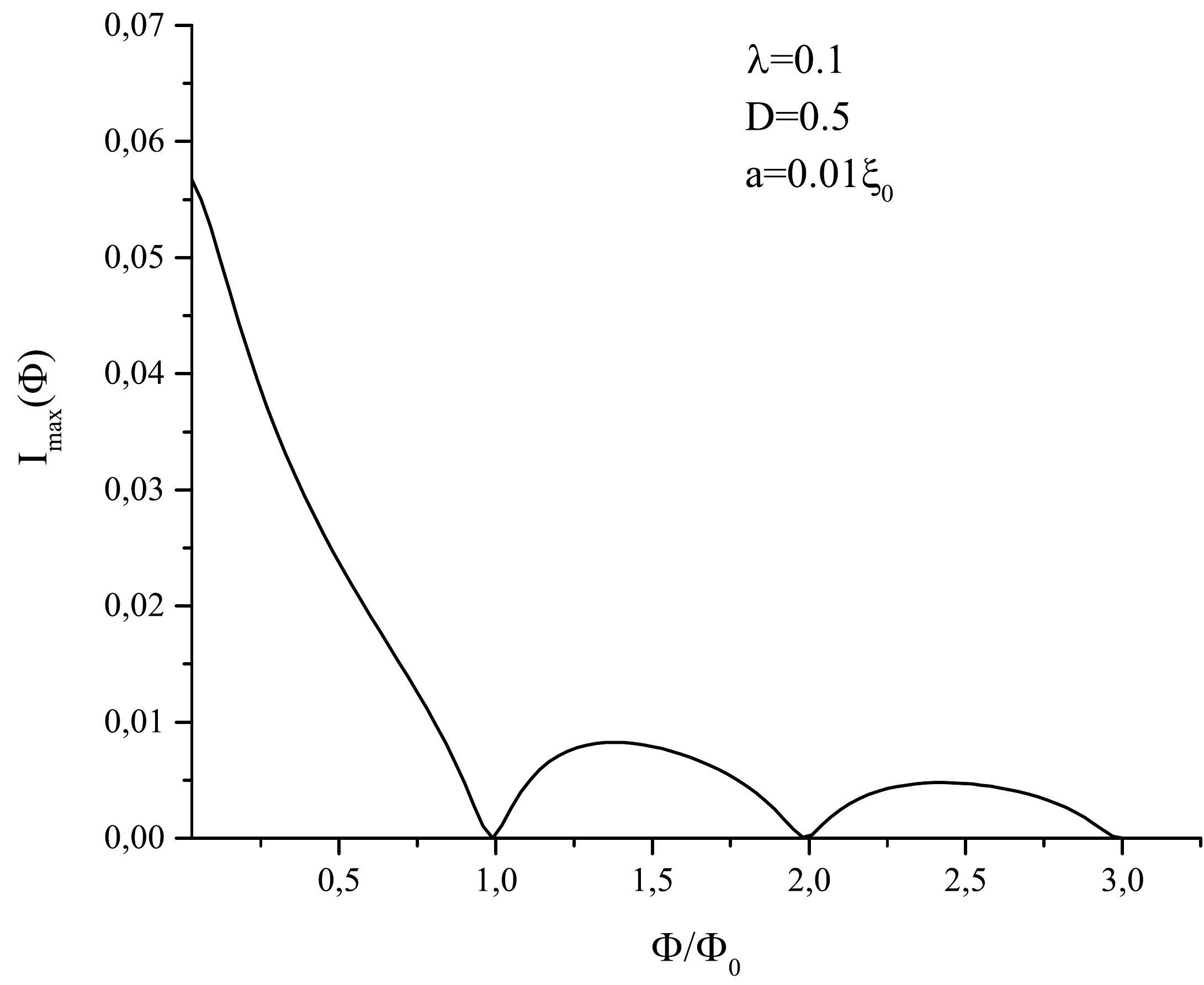}}
	\caption{Dependence of the maximum current on the magnetic flux for different values of electron free path.}
	\label{fig4}
\end{figure}

Figure~\ref{fig4} displays the change of the magnitude and the shape of the maximum current with the change in the electron free path for a given value of the electron transparency coefficient and the thickness of the normal layer. One can see that an increase in the concentration of nonmagnetic impurities leads to a noticeable decrease in the magnitude of the current maxima, while the form of the dependence shows a small decrease in the sensitivity of the junction to the magnetic field. We can also consider numerical comparisons. It is obvious that with a decrease in the electron free path from $\lambda = 1$ to $\lambda = 0.1$, the maximum value $I_{\max } (\Phi )$ decreases by more than 4 times. As for the shape of the dependence $I_{\max } (\Phi )$, for the value of the flow $\Phi / \Phi_0 = 0.4$ at $\lambda = 1$, the maximum current decreases by 2.5 times the maximum possible value, and when $\lambda = 0.1$, the current decrease does not exceed $50\%$.

\section{Conclusions}

In this paper, the behaviour of the superconducting SNINS junction in an external magnetic field near the critical temperature has been investigated. It has been found that the presence of the dielectric layer and the normal region in the junction significantly affect the magnitude of the maximum current and the shape of the current dependence on the magnetic field. The increase in the electron transmission coefficient has an especially noticeable impact on the sensitivity of the maximum current to the change in the magnetic flux, which confirms the possibility of the practical application of such superconducting structures. In addition, the effect of nonmagnetic impurities on the magnitude of the current was demonstrated. It is shown that the change of the electron free path significantly affects the magnitude of the maximum current in a junction, whereas the shape of the maximum current dependence on the magnetic flux through the junction does not show significant changes. We can summarize that with an increase of the electron transmission coefficient and a decrease of the normal layer thickness, when the depairing effects are significant and the current-phase relation is anharmonic, the sensitivity of the junction to the change in the magnetic flux increases. An interesting issue for a further research can be the behaviour of such superconducting junctions in a magnetic field, taking into account the self-induced magnetic field caused by the current passing through the junction.

\subsection*{Acknowledgements}

Present research is supported by the State Fund for the Fundamental Research of Ukraine (award \textnumero F76/123-2017).

\ukrainianpart

\title{Дослідження впливу зовнішнього магнітного поля на максимальний струм SNINS-контактів поблизу критичної температури}

\author{О.Ю. Пастух, В.Є. Сахнюк, А.В. Свідзинський}
\address{Східноєвропейський національний університет імені Лесі Українки,
	просп. Волі, 13, 43000 Луцьк, Україна}

\makeukrtitle

\begin{abstract}
\tolerance=3000%
В роботі досліджено поведінку надпровідних контактів типу SNINS з ангармонійною струм-фазовою залежністю у зовнішньому магнітному полі поблизу критичної температури. Проаналізовано залежність максимального струму від потоку зовнішнього магнітного поля для широкого інтервалу зміни коефіцієнта проходження електронів. Також досліджено, як впливає на чутливість струму до зміни магнітного потоку наявність нормального прошарку довільної товщини в масштабі довжини когерентності та наявність у надпровідних областях немагнітних домішок.
\keywords контакт Джозефсона, максимальний струм, магнітний потік, ефекти розпаровування

\end{abstract}

\end{document}